\documentclass{article}

\usepackage[table]{xcolor}
\usepackage[noadjust]{cite}
\usepackage{capt-of}
\usepackage{amsmath}
\usepackage{amssymb}
\usepackage{braket}
\usepackage{bm}
\usepackage{scrextend} 
\usepackage{mathtools}
\usepackage{graphicx}
\usepackage{stfloats}  
\usepackage{algorithm}
\usepackage{algorithmicx,algpseudocode}
\usepackage{tikz}
\usepackage{amsthm}
\usepackage{cite}
\usepackage[normalem]{ulem}

\usepackage[margin=2cm]{geometry}
\usepackage{newtxtext}
\usepackage{newtxmath}
\usepackage{setspace}
\usepackage[absolute]{textpos}

\usetikzlibrary{automata,positioning}

\makeatletter
\def\ps@IEEEtitlepagestyle{%
	\def\@oddfoot{\mycopyrightnotice}%
	\def\@evenfoot{}%
}
\def\mycopyrightnotice{%
	{\footnotesize  978-1-5386-3406-6/17/\$31.00 \textcopyright2017 Crown\hfill}
	\gdef\mycopyrightnotice{}
}
\makeatother

\makeatletter
\def\thmheadbrackets#1#2#3{%
	\thmname{#1}\thmnumber{\@ifnotempty{#1}{ }\@upn{#2}}%
	\thmnote{ {\the\thm@notefont[#3]}}}
\makeatother



\newtheoremstyle{brakets}
{}
{}
{\itshape}
{}
{\bfseries}
{.}
{ }
{\thmheadbrackets{#1}{#2}{#3}}

\newtheoremstyle{defbrakets}
{}
{}
{\normalfont}
{}
{\bfseries}
{.}
{ }
{\thmheadbrackets{#1}{#2}{#3}}

\newtheoremstyle{defproblem}
{}
{}
{\normalfont}
{}
{\bfseries}
{.}
{ }
{\thmheadbrackets{#1}{#2}{#3}}

\theoremstyle{brakets}
\newtheorem{thm}{Theorem}

\theoremstyle{defbrakets}
\newtheorem{defn}[thm]{Definition}

\theoremstyle{defproblem}
\newtheorem{plm}{Problem}

\algnewcommand\algorithmicforeach{\textbf{for each}}
\algdef{S}[FOR]{ForEach}[1]{\algorithmicforeach\ #1\ \algorithmicdo}

\algnewcommand\algorithmicinput{\textbf{Input:}}
\algnewcommand\INPUT{\item[\algorithmicinput]}

\algnewcommand\algorithmicoutput{\textbf{Output:}}
\algnewcommand\OUTPUT{\item[\algorithmicoutput]}

\let\oldReturn\Return
\renewcommand{\Return}{\State\oldReturn}

\makeatletter
\newcommand{\algrule}[1][.2pt]{\par\vskip.5\baselineskip\hrule height #1\par\vskip.5\baselineskip}
\makeatother

\makeatletter
\let\old@ps@IEEEtitlepagestyle\ps@IEEEtitlepagestyle
\def\confheader#1{%
	\def\ps@IEEEtitlepagestyle{%
		\old@ps@IEEEtitlepagestyle%
		\def\@oddhead{\strut\hfill#1\hfill\strut}%
		\def\@evenhead{\strut\hfill#1\hfill\strut}%
	}%
	\ps@headings%
}
\makeatother

\begin{document}

\title{Automatic Reconfiguration of Untimed Discrete-Event Systems}

\author{M. Macktoobian and W. M. Wonham}
\date{Department of Electrical and Computer Engineering\\
	University of Toronto\\
	Toronto, ON, Canada M5S 3G4\\
	Email: \{matin.macktoobian,wonham\}@scg.utoronto.ca}

\maketitle

\begin{textblock}{14}(1.5,1)
	\noindent\textbf{\color{red}Published in ``2017 14th International Conference on Electrical Engineering, Computing Science and Automatic Control (CCE)''\\ DOI: 10.1109/ICEEE.2017.8108839}
\end{textblock}

\begin{abstract}

This work introduces a general formulation of the
	reconfiguration problem for untimed discrete-event systems (DES),
	which can be treated directly by supervisory control theory (SCT).
	To model the reconfiguration requirements we introduce the concept 
	of \textit{reconfiguration specification} (RS); here \textit{reconfiguration events} (RE) are introduced to force a transition from one system 
	configuration to another. Standard SCT synthesis is employed to 
	obtain a \textit{reconfiguration supervisor} (\textbf{RSUP}) in which designated 
	states serve as the source states for RE.  The reconfiguration 
	problem itself is formulated as that of establishing guaranteed 
	finite reachability of a desired RE source state in \textbf{RSUP} from the current state in \textbf{RSUP} at which a change in configuration is commanded by an external user. The solvability (or otherwise) of 
	this reachability problem is established by backtracking as in
	standard dynamic programming.

\end{abstract}
\doublespacing
\section{Introduction}
Highly complex systems are typically composed of many components. A system configuration (mode) is implemented by a particular subset of the system's components exhibiting a specific functionality. Dynamic reconfiguration is the automatic switching of the current configuration of the system to a new one commanded by the system's external user (e.g., manager of a factory workcell). The importance of reconfiguration, in the case of complex systems, is two-fold. First, the model now embraces multimodal systems, thereby achieving multifunctionality through multimodality. Second, since reconfiguration enables switching from a faulty configuration to a sound one, the overall system's reliability, and its robust response to faults, are each enhanced.
 
There have been extensive investigations into the notion of reconfiguration in the context of different complex systems such as power systems (\cite{abdelaziz2010distribution,gomes2006new,mcdermott1999heuristic}), embedded systems (\cite{voros2009dynamic,van2016model}), hybrid systems (\cite{momayyezan2016integrated,wang2016highly}), and manufacturing systems (\cite{sanderson2016smart}). Petri-net-based modeling of DES has also been employed to realize dynamic reconfiguration tasks. For example, \cite{khalgui2011reconfiguration} proposed a communication protocol to coordinate reconfiguration of software agents by coordination matrices. Many other applications of Petri nets could be cited of dynamic reconfiguration of complex systems (e.g., see \cite{matos2016reconfiguration}).

Moreover, some efforts have been applied to engineer reconfigurable DES. For instance, \cite{zhang2015reconfigurable} recommended a coordinated approach to solve the problem; the technique stressed communication among the subsystems of the DES. It was further assumed that all subsystems are equipped with coordinators to manage the system reconfigurations. The virtual coordinator exchanges messages with subsystems separately, rather than subsystems intercommunicating directly. Whenever a subsystem desires to execute a global reconfiguration, it first requests permission of the coordinator. However, the approach seems not to be efficiently scalable.

Recently, the notion of state attraction \cite{brave1990stabilization} has been used to reconfigure DES by regular specifications. In particular, a coordination automaton (CA) is generated to weakly attract a required reconfiguration. The path from each initial state of CA to its marked state guides a particular supervisor to realize the required reconfiguration \cite{nooruldeen2015state}. However, this approach is unable to reconfigure DES at any arbitrary state, and the synthesized CA is nondeterminstic including more than one initial state.

For the reconfiguration control of DES, SCT \cite{wonham2017supervisory} has already shown 
promise. Recall that in SCT a supervisor is synthesized for given 
formal specifications in such a way as to be \textit{correct by design}. In this spirit \cite{wang2016dynamic} addressed the dynamic rescheduling of a timed DES (or TDES) by use of a timed supervisor. In addition \cite{sampath2008control} proposed a method based on Petri nets and integer programming to 
optimize the cost of reconfiguration. The method uses linear constraints to synthesize deadlock-free supervisors by SCT.

This paper presents an automatic procedure to reconfigure DES by SCT dynamically. To this end, we introduce the notion of reconfiguration specification (RS) to model reconfiguration requirements; this specification is incorporated into the SCT supervisor synthesis in the usual way. The resulting supervisor not only realizes the behavioral specifications of the DES, but also switches between configurations when reconfiguration is commanded by an external system manager. Finally, it is necessary to know whether a desired reconfiguration is achievable from a particular state of the supervisor or not. At that point we apply a backtracking path-finding algorithm based on dynamic programming to collect all forcible paths (if any) that solve the intended reconfiguration problem.

The paper is organized as follows: Section \ref{sec:prel} is devoted to preliminaries. Section \ref{sec:UDES} elaborates on the realization of the automatic dynamic reconfiguration by SCT. The new definition of RS is stated in Section \ref{subsec:spec}. Section \ref{subsec:sup} provides the \textbf{RSUP} synthesis process. The reconfiguration problem is formally defined in Section \ref{subsec:prob}. In Section \ref{subsec:AFC} we describe the collection of forcible backtracking paths;  the formal algorithm is presented Section \ref{subsec:alg}. Section \ref{subsec:exam} illustrates the proposed method with an example, and conclusions are drawn in Section \ref{sec:conc}.
\section{Preliminaries}
\label{sec:prel}
SCT controls DES modeled by the Ramadge-Wonham framework \cite{ramadge1987supervisory,wonham2015supervisory}. A DES is formally represented by a generator, say
	\begin{equation*}
	\textbf{G} = (Q,\Sigma,\delta,q_0,Q_m),
	\end{equation*}
\noindent where $\Sigma = \Sigma_{c} \dot{\cup} \Sigma_{u}$ is a finite alphabet of event labels, partitioned into the \textit{controllable} event labels and the \textit{uncontrollable} ones; $Q$ is the \textit{state set}, assumed finite; $\delta:Q \times \Sigma^{*} \rightarrow Q$ is the \textit{extended partial transition function}; $q_0$ is the \textit{initial state}; and $Q_{m} \subseteq Q$ is the subset of \textit{marked states}. The \textit{closed behavior} and the \textit{marked behavior} of \textbf{G} are the languages
\begin{equation*}
	L(\textbf{G}) := \{s\in \Sigma^{*}|\delta(q_0,s)!\}
\end{equation*}
and 
\begin{equation*}
	L_{m}(\textbf{G}) := \{s\in L(\textbf{G})|\delta(q_0,s) \in Q_{m}\}
\end{equation*}
\noindent Here $\delta(q_0,s)!$ means that $\delta(q_0,s)$ is defined.
A supervisory control function for \textbf{G} is a map $\textbf{V}:L(\textbf{G}) \rightarrow \Gamma$, in which $\Gamma = \{\gamma \in Pwr(\Sigma)|\gamma \supseteq \Sigma_{u}\}$ is the set of \textit{control patterns}. `\textbf{G} under supervision of \textbf{V}' is written as $\textbf{V}/\textbf{G}$. 

Given a sublanguage $M \subseteq L_{m}(\textbf{G})$ we define 
	the \textit{marked behavior} of $\textbf{V}/\textbf{G}$ as $L_{m}(\textbf{V}/\textbf{G}) := L(\textbf{V}/\textbf{G}) \cap M$. $\textbf{V}$ is a 
	\textit{marking nonblocking supervisory control} (MNSC) for the pair $(M,\textbf{G})$ if $ \overline{L_{m}(\textbf{V}/\textbf{G})} = L(\textbf{V}/\textbf{G})$.
In practice, \textbf{V} is implemented by a \textit{supervisor}, which can be taken as a generator \textbf{SUP} that represents the maximally permissive controlled behavior $L_{m}(\textbf{V}/\textbf{G})$ subject to a generator specification, say \textbf{SPEC}; we denote this computation by \textbf{SUP} = \textbf{supcon} (\textbf{G},\textbf{SPEC}). For details see, e.g., \cite{wonham2017supervisory}.
\begin{figure}[t]
	\centering\includegraphics[scale=0.9]{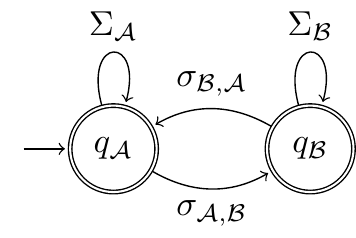}
	\caption{Generator of a binary RS}
	\label{fig:many-many}
\end{figure}
\section{DES Reconfiguration}
\label{sec:UDES}
\begin{figure*}
	\includegraphics[width=\textwidth]{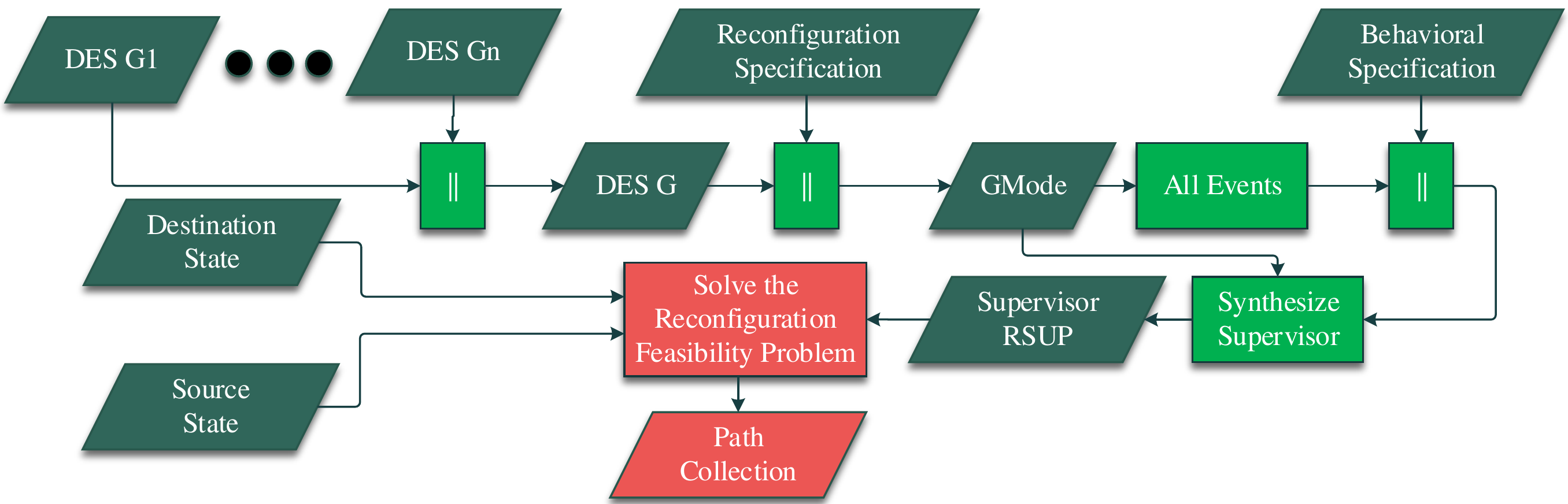}
	\caption{DES reconfiguration process}
	\label{fig:UDES}
\end{figure*}
\subsection{Reconfiguration Specification}
\label{subsec:spec}
A reconfiguration requirement is a transition request to transfer the system from one configuration to another. The set of reconfiguration requirements associated with a DES is represented by the RS. As the first step, the following definition clarifies the notion of configuration (mode). 
	\begin{defn}[Configuration (Mode)]
	Let \textbf{G} be a DES constructed by (synchronous product of) the collection of component sets $C_{\textbf{G}}=\{C_{\textbf{G}}^{\alpha}|\alpha \in \mathcal{I}\}$, where $\mathcal{I}$ is an index set. A \textit{configuration (mode)} of \textbf{G} is a component set $C^{0}_{\textbf{G}} \subseteq C_{\textbf{G}}$ such that \textbf{G} operates as the corresponding synchronous subproduct, when the elements of $C^{0}_{\textbf{G}}$ and $C_{\textbf{G}}\setminus C^{0}_{\textbf{G}}$ are respectively active and inactive.
\end{defn}
The source and destination configurations, corresponding to a reconfiguration task, are connected through a reconfiguration event (RE) which specifies a transition from the former to the latter.
\begin{defn}[RE]
	\label{def:rec_eve}
	Given DES \textbf{G} with component set $C_{\textbf{G}}$, let $C^{\mathcal{A}}_{\textbf{G}},C^{\mathcal{B}}_{\textbf{G}} \subset C_{\textbf{G}}$ be two configurations of \textbf{G}, where $\mathcal{A}$ and $\mathcal{B}$ are the component sets which must be deactivated and activated respectively in the course of reconfiguration; then, RE $\sigma_{\mathcal{A},\mathcal{B}}$ must occur to trigger the reconfiguration task. We formalize this task as follows.
\end{defn}
\begin{defn}[RS]
	\label{def:rec_spec}
	For a configuration $C_{\textbf{G}}^{\mathcal{I}}$ of $\textbf{G}$, with components $\{\textbf{G}_{a} | a \in \mathcal{I}\}$, let $\Sigma_{\mathcal{I}}$ denote the set of exactly the event (labels) appearing in	the corresponding components, i.e.
	\begin{equation*}
		\Sigma_{\mathcal{I}} := \bigcup\{\Sigma_{a} | a \in \mathcal{I}\}
	\end{equation*}
	For clarity we first consider the case of just two configurations, 
	indexed by $\mathcal{A}$ and $\mathcal{B}$, and assume that reconfiguration requires 
	switching either from $C_{\textbf{G}}^{\mathcal{A}}$ to $C_{\textbf{G}}^{\mathcal{B}}$, with the system 
	$\textbf{G}$ initially in mode $C_{\textbf{G}}^{\mathcal{A}}$.  As in Definition \ref{def:rec_eve}, bring in the corresponding switching events $\sigma_{\mathcal{A},\mathcal{B}}$ and $\sigma_{\mathcal{B},\mathcal{A}}$.  We define the RS as the DES
	\begin{equation*}
	E_\mathcal{R} := (Q_\mathcal{R},\Sigma_\mathcal{R}, \delta_\mathcal{R},{q_{0}}_\mathcal{R},{Q_m}_\mathcal{R})
	\end{equation*}
	where
	\begin{itemize}
		\item $Q_\mathcal{R} := \{q_{\mathcal{A}}, q_{\mathcal{B}}\}$
		
		\item $\Sigma_\mathcal{R}:= \Sigma_{\mathcal{A}} \cup \Sigma_{\mathcal{B}} \dot{\cup} \{\sigma_{\mathcal{A},\mathcal{B}}, \sigma_{\mathcal{B},\mathcal{A}}\}$\\ Note that $\Sigma_{\mathcal{A}} \cup \Sigma_{\mathcal{B}}$ is disjoint from $\{\sigma_{\mathcal{A},\mathcal{B}}, \sigma_{\mathcal{B},\mathcal{A}}\}$.
		\item \begin{equation*}
		\delta_\mathcal{R}(q_\mathcal{A},\sigma):=
		\begin{cases}
		q_{\mathcal{A}} ~~~\text{if}~~~ \sigma \in \Sigma_{\mathcal{A}},\\
		q_{\mathcal{B}} ~~~\text{if}~~~ \sigma = \sigma_{\mathcal{A},\mathcal{B}},
		\end{cases}
		\end{equation*}
		\begin{equation*}
		\delta_\mathcal{R}(q_\mathcal{B},\sigma):=
		\begin{cases}
		q_{\mathcal{B}} ~~~\text{if}~~~ \sigma \in \Sigma_{\mathcal{B}},\\
		q_{\mathcal{A}} ~~~\text{if}~~~ \sigma = \sigma_{\mathcal{B},\mathcal{A}},
		\end{cases}
		\end{equation*}
		
		\item ${q_{0}}_\mathcal{R} := q_{\mathcal{A}}$
		
		\item ${Q_{m}}_\mathcal{R} := \{q_{\mathcal{A}},q_\mathcal{B}\}$
	\end{itemize}
\end{defn}
$E_\mathcal{R}$ is depicted in Fig. \ref{fig:many-many}.

It is obvious that, with some notational elaboration, this definition 
	of $E_\mathcal{R}$ could be extended to an arbitrary number of distinct configurations and pairwise switchings among them, but we refrain from doing so here.
\subsection{Reconfiguring Supervisor Synthesis}
\label{subsec:sup}
In general SCT synchronizes the plant DES \textbf{G} with its associated behavioral specification prior to synthesizing the supervisor which controls it. To synthesize \textbf{RSUP} we provide in this section an extended supervisor synthesis procedure in which the RS is taken into account.

The main idea is to treat the RS on the same footing as any other behavioral specification. The process is illustrated in Fig. \ref{fig:UDES}.

We start by synchronizing all the components associated with  \textbf{G}; here the operator `$\parallel$' denotes \textit{synchronous product} \cite{tct2017supervisory}. Then, \textbf{G} is synchronized with the RS $E_\mathcal{R}$ leading to the multimodal version of \textbf{G}, say \textbf{GMode} as follows.
\begin{equation*}
\textbf{GMode} := (||_{i=1}^{n} \textbf{G}_{i}) || E_\mathcal{R}
\end{equation*}
Next, the global system specification is defined by the synchronization of the behavioral specification $E$ with all events of \textbf{G}Mode. Finally, the reconfiguration supervisor is computed as:
\begin{equation*}
\textbf{RSUP} := \textbf{supcon} (\textbf{GMode}, [\textbf{allevents}(\textbf{GMode})||E])
\end{equation*}
\noindent where the operator $\textbf{allevents}$ returns a marked one-state DES selflooped with all the events of the argument. For details see, e.g., \cite{wonham2017supervisory}.

We emphasize that \textbf{RSUP} enforces both the behavioral and reconfiguration requirements of \textbf{G}.

In the next section we complete the design process by presenting an algorithm to solve the final reconfiguration problem.
\section{Reconfiguration Solvability}
\label{sec:solv}

We construct a backtracking algorithm that collects all forcible paths reaching a suitable target state of \textbf{RSUP} from an arbitrary current state. To this end, we first present the formal definition of the reconfiguration problem, then obtain the conditions for backtracking forcibility, and finally propose an algorithm to solve the problem. 
\subsection{Problem Statement}
\label{subsec:prob}

Especially when it is the intended response to some unexpected failure or emergency, reconfiguration could possibly be commanded from an arbitrary state in \textbf{RSUP} of the current (running) configuration. Thus the system must be forced onto a path or set of paths in the \textbf{RSUP} state space leading from the current state to one at which the desired reconfiguration event is defined. The question is then whether or not such a path (set) always exists and can be executed in timely fashion.
We thus consider
\begin{plm}[Reconfiguration Feasibility]
	Denote by $q_c$ the state where 
	\textbf{RSUP} currently resides, and let $q_r$ be a state at which a desired 
	reconfiguration event, say $\sigma_r$, is defined; namely 
	$\delta_\mathcal{R}(q_r,\sigma_r)!$  Subject to an appropriate specification of 
	\textit{forcibility}, determine the forcible path (set) from $q_c$ to $q_r$.
\end{plm}
We shall declare a path (set) \textit{forcible} if it can be enforced using the standard SCT ``technology" of controllable event disablement, with
or without the possible additional use of preemption by a forcible 
event (\hspace*{-1mm}\cite{wonham2017supervisory}, Sect. 3.8).

The following section describes how to obtain a forcible path (set).
\subsection{Backtracking Forcibility}
\label{subsec:AFC}

The algorithm is critically governed by the conditions which define  forcibility of each backtracking step; if a backtracking step is forcible, then its event is appended to the current path. 

\begin{figure}[t]
		\centering\includegraphics[scale=0.85]{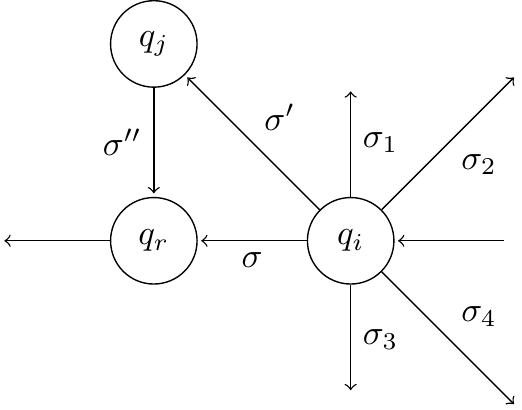}
		\caption{The hypothetical backtracking scenario}
		\label{fig:fig3}
	\end{figure}

We explain the backtracking forcibility conditions (BFCs) by a hypothetical search space as depicted in Fig. \ref{fig:fig3}. We assess the forcibility of the backtracking from $q_r$ to $q_i$. In other words, we plan to find whether $\sigma$ contributes to an forcible path (from $q_i$ to $q_r$) or not. 

Let $\Sigma^{i} = \{\sigma_1,\sigma_2,\sigma_3,\sigma_4,\sigma'\}$ be the set of defined events at $q_i$, competing with $\sigma$ to occur. Let also $\Sigma_{i}=\{\sigma\}$ be the defined events transitioning from $q_i$ to $q_r$. If any of the following four conditions hold, then backtracking to $q_i$ is permissible, and $\sigma$ is a solution to this problem:

\textbf{BFC-1}: $\Sigma^{i} = \varnothing$\\
	There is no event competing with $\sigma$ to occur at $q_i$. In other words, $\sigma$ is the exclusive event which can occur at $q_i$.
	
\textbf{BFC-2}: $(\forall \sigma' \in \Sigma^{i})~ \sigma' \in \Sigma_{c}$\\
	If all of the $\Sigma^{i}$ events are controllable, then they can be disabled; therefore, $\sigma$ can occur at $q_i$.
	
\textbf{BFC-3}: $[(\nexists \sigma' \in \Sigma_{i}) ~\text{pr}(\sigma', \sigma)] ~~\wedge~~ [(\forall \sigma' \in \Sigma^{i}_{u})(\exists \sigma'' \in  \Sigma^{i})] ~\text{pr}(\sigma'',\sigma')$\\
	$\Sigma^{i}$ is composed of both controllable and uncontrollable events. The first subformula says no competing event should be able to preempt $\sigma$. The second subformula asserts that each uncontrollable event must be preemptable by at least another event. Note that the binary relation $\text{pr}(\sigma', \sigma'')$ holds if the forcible event $\sigma'$ can preempt $\sigma''$ (see, section 3.9 of \cite{wonham2017supervisory}). Furthermore, $\Sigma_{u}^{i}$ and $\Sigma_{c}^{i}$ represent the largest uncontrollable and controllable subsets of $\Sigma^{i}$, respectively, where $\Sigma^{i} = \Sigma_{c}^{i} \dot{\cup} \Sigma_{u}^{i}$.
	
\textbf{BFC-4}: $(\exists \sigma' \in \Sigma_{u}^{i})(\forall \sigma'' \in \Sigma_{j} \cup \Sigma^{j})~~[\delta(q_i, \sigma') = q_j \wedge \sigma'' \in \Sigma_{c}$], where $\Sigma_{j}$ contains the event(s) transitioning from $q_j$ and to $q_r$, and $\Sigma^{j}$ contains the event(s) transitioning from $q_j$ to the other states.

	This condition determines the succession of the uncontrollable events leading to a state containing only controllable events; since the controllable events can be disabled, both $\sigma$ and $\sigma'\sigma''$ are forcible paths.

We employ the foregoing conditions, as decision-making criteria, to solve the reconfiguration problem by a backtracking algorithm in the next section.
\subsection{Reconfiguration Solvability Algorithm}
\label{subsec:alg}
Admissible paths can be collected by the Algorithm \ref{alg:RCA}.
The inputs are the target state, from which the backtracking starts; the source state, which is the algorithm's goal to be visited; the state list of the supervisor, which is the search space of the problem including the target and source state; and a temporary list of incomplete paths initialized to empty, in which the potential solutions are stored, corresponding to each instantiation of the algorithm. The output is the collection of the found paths.

The algorithm keeps the track of the visited states and events. If it visits a state that is the destination of more than one event, the algorithm checks inductively the solvability of those branches.

\begin{algorithm}[t]
	\caption{Path Collector Algorithm ($\mathcal{PCA}$)}
	\label{alg:RCA}
	\begin{algorithmic}[1] 
		\INPUT 
		\Statex $q_r$ \Comment Target state
		\Statex $q_s$ \Comment Source state
		\Statex $Q_T$  \Comment State list
		\Statex $\mathcal{P}_{\text{temp}}$  \Comment Temporary list of  an incomplete path
		\OUTPUT 
		\Statex $\mathcal{P}$ \Comment Path collection
		\algrule[1pt]
		\If{$Q_T = \varnothing$}
		\Return $\mathcal{P}$ \Comment All instances of $\mathcal{PCA}$ are terminated.
		\Else {}
		\State $\mathcal{Q} \leftarrow \{q|\delta(q,\sigma)=q_r\}$
		\ForEach {$q_i \in \mathcal{Q}$}
		\State $\Sigma_{i} \leftarrow \{\sigma|\delta(q_i,\sigma)=q_r$\}
		\State $\Sigma^{i} \leftarrow \{\sigma|\delta(q_i,\sigma)!\} - \Sigma_{i}$
		\If {$q_i = q_s$}
		\State $\mathcal{P} \leftarrow \text{append } \Sigma
		_{i} \text{ to } \mathcal{P}_{\text{temp}}$
		\State $Q_T \leftarrow Q_T - \{q_i\}$
		\Else {}
		\ForEach{$\sigma \in \Sigma_{i}$}
		\If{(BFC-1 \hspace*{-2mm} $\parallel$ \hspace*{-2mm} BFC-2 \hspace*{-2mm} $\parallel$ \hspace*{-2mm} BFC-3 \hspace*{-2mm} $\parallel$ \hspace*{-2mm} BFC-4)
		}
		\State $\mathcal{P}_{\text{temp}} \leftarrow \text{append } \sigma \text{ to } \mathcal{P}_{\text{temp}}$
		\State $Q_T \leftarrow Q_T - \{q_i\}$
		\Return $\mathcal{PCA}(q_i,q_s,Q_T,\mathcal{P}_{\text{temp}})$
		\EndIf
		\EndFor
		\EndIf
		\EndFor
		\EndIf
	\end{algorithmic}
\end{algorithm}

{\itshape Line 1} defines the terminal case; if all states are checked, then the path collection will be delivered. 

{\itshape Line 4} gathers the states that have transitions to the current target state. The defined events at each found state are stored in two variables: variable $\Sigma_{i}$ which contains the event(s) transitioning from the found state $q_i$ to the current state ({\itshape Line 6}); and variable $\Sigma^{i}$ which stores the event(s) transitioning from $q_i$ to the other states ({\itshape Line 7}).

If $q_i$ is the source state ({\itshape Line 8}), then a path is found. Hence, the recent found event is attached to the temporary solution, and the result is stored in the path collection ({\itshape Line 9}). Moreover, $q_i$ is removed from the global list of unvisited states ({\itshape Line 10}). Otherwise, BFCs are checked with respect to the elements of $\Sigma_{i}$ and $\Sigma^{i}$ ({\itshape Line 13}). If any of these conditions holds, then the backtracking from the current state to $q_i$ is authorized, thereby appending $\sigma$ to the temporary solution variable ({\itshape Line 14}). Then, the global variable list is updated by removing $q_i$ ({\itshape Line 15}), and a new instantiation of the algorithm is called according to the updated current state ({\itshape Line 16}). Note that both $Q_T$ and $\mathcal{P}$ are global variables that can be read and modified by any instantiation of $\mathcal{PCA}$.

Finally, Algorithm \ref{alg:RCA} can be incorporated appropriately into the final solvability checker function, i.e., Algorithm \ref{alg:RSA}. First, the state list is initialized to the unvisited states; thus, the target state is initially excluded ({\itshape Line 1}). Then, all of the found paths by instantiations of $\mathcal{PCA}$ are gathered ({\itshape Line 2}). The fourth argument of $\mathcal{PCA}$ represents the found path that is initially empty. As already asserted, each successful backtracking step appends the found event to its incomplete path. At the end, the decision is made based on the path collection ({\itshape Lines 3-7}).\begin{algorithm}[t]
	\caption{Reconfiguration Solvability Algorithm ($\mathcal{RSA}$)}
	\label{alg:RSA}
	\begin{algorithmic}[1]
		\INPUT 
		\Statex $q_r$ \Comment Target state
		\Statex $q_s$ \Comment Source state
		\Statex $\textbf{RSUP}$  \Comment Supervisor
		\OUTPUT 
		\Statex solvability
		\algrule[1pt]
		\State $Q_T \leftarrow Q_{\textbf{RSUP}} - \{q_r\}$
		\State $\mathcal{P} \leftarrow \mathcal{PCA}(q_r,q_s,Q_T,\varnothing)$
		\If{$\mathcal{P} \neq \varnothing$}
		\Return The reconfiguration problem \textbf{IS} solvable by \hspace*{1.62cm} any path of $\mathcal{P}$.
		\Else {}
		\Return The reconfiguration problem \textbf{IS NOT} solvable. 
		\EndIf
	\end{algorithmic}
\end{algorithm}
\section{Example}
\label{subsec:exam}
We provide an example to illustrate a configuration problem associated with a DES.

\begin{figure}[t]
	\centering\includegraphics[scale=1.4]{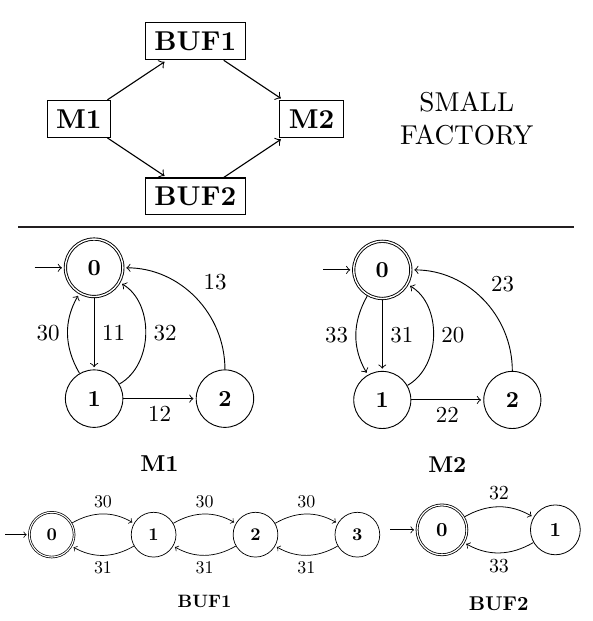}
	\caption{The schematic of SMALL FACTORY}
	\label{fig:examp}
\end{figure}

\begin{figure*}[bt]
	\begin{minipage}{0.54\linewidth}
		\centering
		\includegraphics[scale=1.2]{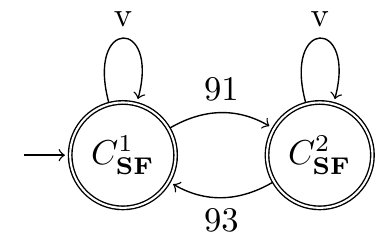}
		\caption{The RS corresponding to SMALL FACTORY\newline(the label v represents the event set $\{11,12,13,20,22,23,30,31,32,33\}$)}
		\label{fig:rec}
	\end{minipage}%
	\begin{minipage}{0.46\linewidth}
		\captionof{table}{Descriptions of SMALL FACTORY events}
		\label{tbl:events} 
		\begin{tabular}{|c||l|}
			\hline
			Event & Description\\
			\hline
			11 & \textbf{M1} takes a workpiece.\\
			20 & \textbf{M2} successfully completes processing a workpiece.\\
			12~/~22 & Breakdown occurs in \textbf{M1}~/~\textbf{M2}.\\
			13~/~23 & \textbf{M1}~/~\textbf{M2} is repaired.\\
			30~/~32 & \textbf{M1} increments \textbf{BUF1}~/~\textbf{BUF2}.\\
			31~/~33 & \textbf{M2} decrements \textbf{BUF1}~/~\textbf{BUF2}.\\
			\hline
			91 & The reconfiguration $C^{1}_{\textbf{G}} \rightarrow C^{2}_{\textbf{G}}$ is requested.\\
			93 & The reconfiguration $C^{2}_{\textbf{G}} \rightarrow C^{1}_{\textbf{G}}$ is requested.\\
			\hline
		\end{tabular}
	\end{minipage}
\end{figure*}

Figure \ref{fig:examp} depicts SMALL FACTORY (\textbf{SF}) with two 3-state machines \textbf{M1} and \textbf{M2}, and buffers \textbf{BUF1} and \textbf{BUF2} with three slots and one slot, respectively. Assume that the machines can work with either of the buffers. In other words, \textbf{SF} has two modes: $C^{1}_{\textbf{SF}} = \{\textbf{M1},\textbf{M2},\textbf{BUF1}\}$, and $C^{2}_{\textbf{SF}} = \{\textbf{M1},\textbf{M2},\textbf{BUF2}\}$. Moreover, the events are described in Table \ref{tbl:events}. Given the required RS as Fig. \ref{fig:rec}, the synthesized \textbf{RSUP} is controllable including 78 states and 270 events.

Assume that SF resides currently at [\textbf{41}] \footnote{In this section, symbols $[q]$ and $\Braket{\sigma}$ represent state $q$ and event $\sigma$, respectively.} and a reconfiguration is requested to switch from $C^{1}_{\textbf{SF}}$ to $C^{2}_{\textbf{SF}}$. Let [\textbf{9}] be the destination state, including $\braket{91}$ where the reconfiguration event $\braket{91}$ is defined.

\begin{figure*}[b]
	\includegraphics[width=\textwidth]{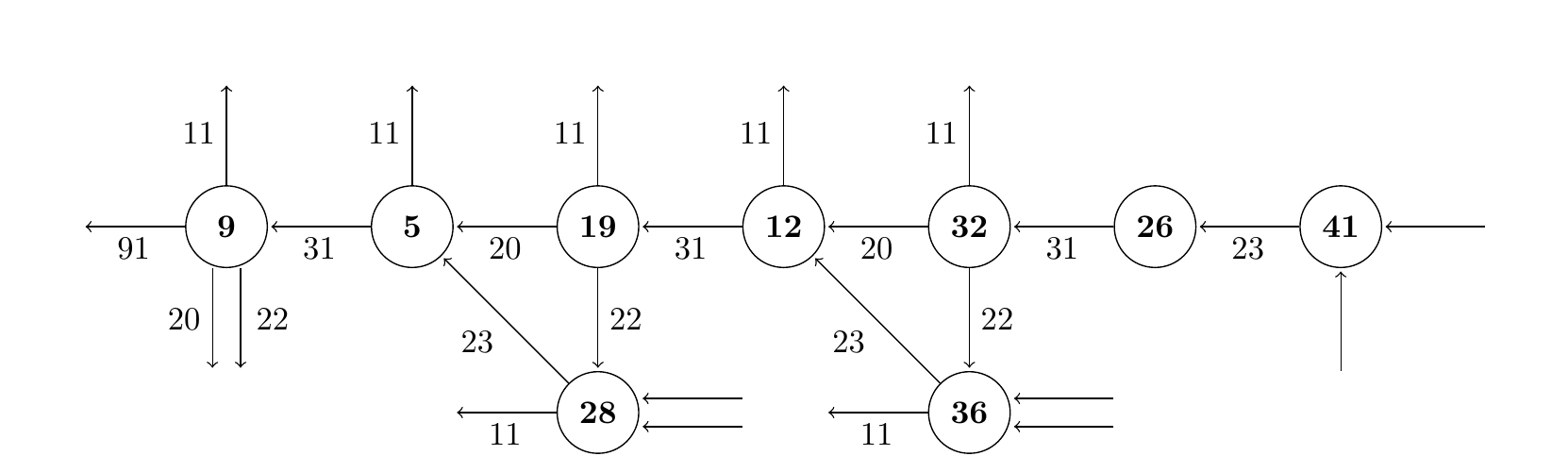}
	\caption{The forcible backtracking goes from [\textbf{9}] back to [\textbf{41}]}
	\label{fig:ex}
\end{figure*}

Consequently, $\mathcal{RSA}$ backtracks from the [\textbf{9}] to [\textbf{41}] to find all forcible paths. The shortest path is $\braket{23,31,20,31,20,31}$, as depicted in Fig. \ref{fig:ex}. The path computation is analyzed as follows.  
\begin{itemize}
	\item Backtracking [\textbf{9}] to [\textbf{5}]: according to BFC-2, backtracking is authorized; because $\braket{11}$ can be disabled.
	
	\item Backtracking [\textbf{5}] to [\textbf{19}]: $\braket{11}$ can be safely disabled, but $\braket{22}$ may occur sooner than $\braket{20}$, thereby preempting it according to BFC-3. According to the BFC-4, $\braket{23}$ can be disabled, so the occurrence of $\braket{22}$ is handled and blocked by the aforesaid disablement. Therefore, both paths $\braket{20}$ and $\braket{22,23}$ are qualified paths. 
	
	\item Backtracking [\textbf{19}] to [\textbf{12}]: the same as backtracking [\textbf{9}] to [\textbf{5}].
	
	\item Backtracking [\textbf{12}] to [\textbf{32}]: the same as backtracking [\textbf{5}] to [\textbf{19}].

	\item Backtracking [\textbf{32}] to [\textbf{26}]: according to BFC-1, backtracking is authorized; because there is no event competing with $\braket{31}$ to occur.
	
	\item Backtracking [\textbf{26}] to [\textbf{41}]: the same as backtracking [\textbf{32}] to [\textbf{26}], considering $\braket{23}$ instead of $\braket{31}$.
\end{itemize}

The results demonstrate that the reconfiguration problem is solvable, and the supervisor can randomly select any path from the forcible path collection to realize the reconfiguration.

It is important to note, finally, that at state [\textbf{9}], which serves as source state for the RE $\braket{91}$  the controllable event $\braket{11}$ must be disabled, and $\braket{91}$ must be forced to preempt the uncontrollable events $\braket{20,22}$. This is acceptable on physical grounds, inasmuch as $\braket{11}$ (\textbf{M1} takes a new workpiece) is controllable, and $\braket{20,22}$ can only occur while \textbf{M2} is working, so that preemption can be plausibly assumed to take place in the working time delay interval.

Note that the 2-way RS may result in blocking in case of retaining the possibility of a reconfiguration from $C^{2}_{\textbf{SF}}$ back to $C^{1}_{\textbf{SF}}$, since disabling $\braket{91}$ ends up with the disablement of another reconfiguration. We may address two approaches to solving this issue: first, $C^{1}_{\textbf{SF}}$ can be considered as the initial state of RS, thereby solving the second reconfiguration problem independently of the first one; alternatively, a 1-way RS can be used for each reconfiguration.

\section{Conclusion}
\label{sec:conc}
This paper proposes a solution to the reconfiguration problem for DES. To this purpose, we first model reconfiguration requirements of the DES by RS. Then, SCT synthesizes \textbf{RSUP}, considering the RS. Finally, forcible paths are computed by a dynamic-programming-based algorithm, so it can check the solvability of a user-initiated reconfiguration request. The algorithm in fact backtracks from a desired reconfiguration event source state to the current state in \textbf{RSUP}. Guaranteed finite reachability of the source state is confirmed (or otherwise) by the construction of the forcible path set.
\section*{Acknowledgment}
This work is supported by the Natural Sciences and Engineering Research Council (NSERC), Grant Number DG-480599. 

\nocite{*}
\bibliographystyle{IEEEtran}
\bibliography{references}{}

\end{document}